# Plasma Stratification in Radio-Frequency Discharges in Argon Gas

Vladimir I Kolobov,[1,2,*] Robert R Arslanbekov,[1] Dmitry Levko,[1] and Valery A Godyak[3]

[1]*CFD Research Corporation, Huntsville, AL 35806, USA*

[2]*Center for Space Plasma and Aeronomic Research, University of Alabama in Huntsville, Huntsville, AL 35899, USA*

[3]*RF Plasma Consulting, Brookline, MA 02446, USA*

We conducted experimental studies and computer simulations of standing striations in capacitive coupled plasma in Argon gas. Standing striations were observed at frequencies 3.6, 8.4 and 19.0 MHz, in a pressure range 0.05-10 Torr, tube radius R=1.1 cm, for a certain range of discharge currents (plasma densities). Numerical simulations revealed similar nature of standing striations in CCP and moving striations in DC discharges under similar discharge conditions. Comparison of computer simulations with experimental observations helped clarify the nature of these striations. The non-linear dependence of the ionization rate on electron density is shown to be the main underlying mechanism of the stratification phenomena.

Plasma stratification often occurs in laboratory devices and has been observed in space plasmas [1]. The most studied are striations in direct current (DC) glow discharges [2]. Moving striations are usually observed in noble gases, while standing striations are typical for molecular gases. The nature of moving striations (also called ionization waves) in noble gases is relatively well understood [3,4]. Plasma stratification in molecular gases remains poorly understood.

Standing striations have been also observed in radio-frequency (RF) discharges [5]. In Capacitively Coupled Plasma (CCP) created between two wires wrapped around a long dielectric tube, standing striations were observed in Argon over a certain range of frequencies and gas pressures [6,7]. Their wavelength is proportional to the tube radius and decreases with increasing gas pressure. The wavelength varies discontinuously with changing the inter-electrode distance in such a way that an integer number of standing waves always forms between the electrodes. The wavelength weakly depends on driving frequency in the range from 3 MHz to 60 MHz. Although discussions of surface-wave and kinetic effects can be found in the literature, no satisfactory explanations of these striations have been proposed so far. Standing striations have been also observed in Inductively Coupled Plasma (ICP) for certain geometries and discharge conditions [8,9].

Striations appeared also in Particle-in-Cell (PIC) simulations of plasmas in different gases and gas mixtures [10,11,12]. Plasma stratification, as an example of self-organization at the kinetic level, remains a great challenge for the plasma science and physics of gas discharges [13].

Previously, standing striations in argon CCP have been observed in a range of frequencies 3-80 MHz and gas pressure 2 Torr, in a tube with radius $R$=1.25 cm [7]. By changing the driving frequency and the distance $L$ between the electrodes, the researchers were able to change the number of striations. However, the number of striations per unit length remained constant at different interelectrode distances. For driving frequencies of 70, 29, 6, 4.2, and 3.7 MHz, they observed 8, 9, 10, 11, and 12 striations at $L$=30 cm. Light emission oscillated at the second harmonic of the applied voltage. The nature of the striations remained unclear.

In the present paper, we describe experimental studies and computer simulations of standing striations in Argon CCP. Experiments confirmed that standing striations exist in Argon CCP at frequencies 3.6, 8.4 and 19.0 MHz, in a pressure range 0.05-10 Torr, for a certain range of discharge currents (plasma densities). Comparison of the computer simulations with the experimental observations help clarify the nature of these striations.

Nowadays, it is well-known that plasma of DC discharges in noble gases is stratified over a wide range of gas pressures and discharge currents. Several types of striations exist depending on specific values of $pR$ and $i/R$ (here, $p$ is the gas pressure, and $i$ is the discharge current). A specific type of striations is observed at low pressures and high currents, near the so-called Pupp boundary in the ($pR$, $i/R$) diagram, which separates the stratified and non-stratified plasma states on the ($pR$, $i/R$) map [3]. The amplitude of moving striations decreases gradually by crossing the Pupp boundary with increasing discharge current. It is now confirmed that the main mechanism of plasma stratification under these conditions is the nonlinear dependence of the ionization rate on the electron density, whereas gas heating plays a secondary role. Under these conditions, the plasma is partially ionized and collisional; the charged particles are generated via direct or stepwise electron-impact ionization, and either lost at the wall by the ambipolar diffusion

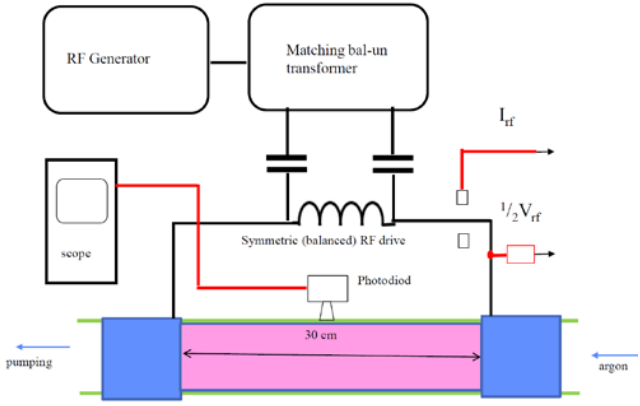

FIG. 1. Experimental device for studies of plasma stratification in RF discharges. The purpose of the gas flow is to prevent any gas contamination from the discharge tube wall

(at low *pR*) or recombine in the volume (at high *pR*).

Our experiments were conducted using a long Pyrex tube with the internal radius $R = 1.1$ cm, the external 5 cm wide foil strip RF electrodes with the inter-electrode distance $L = 30$ cm. Figure 1 shows schematic of the experimental device for studies of plasma stratification in RF discharges. Particular attention was paid to the symmetric powering of both electrodes and the gas purity. We have observed that even small voltage asymmetry can result in slow motion of striations in the axial direction visible by the naked eye.

Experiments were conducted at frequencies 3.6, 8.4, and 19 MHz. Figure 2 shows typical photographs of light emission obtained at 3.6 MHz for different gas pressures. For a given pressure, striations appeared for a certain range of discharge currents (plasma densities). Most pronounced striations were observed for 52 mA at 0.05 Torr, 55 mA for 0.1 Torr, 50-80 mA for 0.2 Torr, and 114 mA for 0.5 Torr (the current values are the root mean square (rms) values). The number of striations does not depend on the RF frequency, but changes slightly with the discharge current. For 19 MHz, weak striations exist only over a narrow range of discharge currents (around 20 mA).

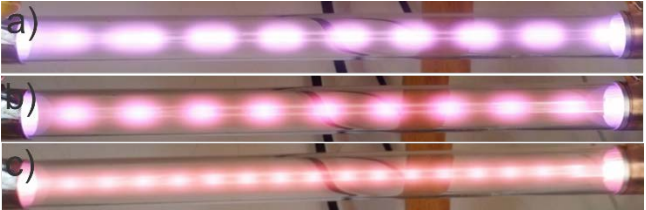

FIG. 2. Photos of standing striations in CCP at 3.6 MHz for pressure (current) a) 0.1 (28), b) 1 (48) and c) 10 (38) Torr (mA).

To explain the observed standing striations in CCP, we apply the previously developed model of plasma striations in DC discharges [3]. The model includes the balance equation for the plasma density and the current with the ionization rate as a strong function of the electron temperature and electron density. The nonlinear dependence of the ionization rate on electron density has proven to be the main cause of constriction and stratification of DC discharges in noble gases near the Pupp boundary.

Electron thermal conductivity is critical for the striation analysis. It is taken into account by assuming that the ionization rate is a function of local electron temperature. The basic set of equations is the following:

$$\nabla \cdot \mathbf{\Gamma} = 0 \qquad (1)$$

$$\frac{\partial n}{\partial t} - \nabla \cdot D_a \nabla n = Z \qquad (2)$$

$$\frac{\partial T_e}{\partial t} - \frac{1}{n_e}\nabla \cdot \chi \nabla T_e = \frac{\mathbf{j}_e \cdot \mathbf{E}}{n_e} - H \qquad (3)$$

The total particle flux $\mathbf{\Gamma}$ is the sum of the electron and ion fluxes, which contain the drift and diffusion components, $n$ is the plasma density and $D_a$ denotes the ambipolar diffusion coefficient. The ionization rate, $Z = n_e v_i(T_e, n_e)$, is a function of the mean electron energy (temperature) and the electron density. The latter is one of the main causes of plasma stratification under the conditions of interest. The electron temperature is defined by the balance between the Joule heating and the energy loss in collisions with neutrals and the generation of new electrons in ionization events described by the $H$ term in Eq. (3). The electron thermal conductivity $\chi = 5n_e D_e / 3$, is proportional to the electron diffusion coefficient $D_e$.

The linear analysis of the discharge stratification was performed using a one-dimensional model with the radial loss of particles to the wall in Eq. (2) approximated as $n / \tau_a$, where $\tau_a = \Lambda^2 / D_a$ is the ambipolar diffusion time, and $\Lambda$ is an effective transverse discharge dimension. Introducing small perturbations of plasma parameters in the form $X = X_0\left(1 + \tilde{X}\exp\left(i(kx - \Omega t)\right)\right)$, where $k = 2\pi / \lambda$ is the wave vector, $\lambda$ is the striation wavelength, and $\Omega = \omega + i\gamma$ is the complex wave frequency, we find the amplification coefficient [3]:

$$\gamma(k) = -D_a k^2 - \frac{1}{\tau_a}\left[\frac{Z_T}{(k\lambda_T)^2} - Z_n\right], \qquad (4)$$

where $\lambda_T = T_e / (eE_0)$, $Z_T = \partial \ln Z / \partial \ln T_e$, $Z_n = \partial \ln Z / \partial \ln n$. The values of electron temperature $T_e$ and the electric field $E_0$ correspond to the solution for the non-stratified positive column plasma. The decrement $\gamma(k)$ has a maximum at some $k = k_0$ with a maximum value $\gamma(k_0) = \frac{Z_n}{\tau_a} - 2D_a k_0^2$. The origin of this maximum has a simple explanation. Short waves decay due to the ambipolar diffusion (the first term in Eq. (4)). Long waves decay because for $k\lambda_T \ll 1$ perturbations of the electric field are opposite in phase with respect to the perturbations of the plasma density ( $\tilde{E} = -\tilde{n}(1 + ik\lambda_T)$ from Eq. (1)). The maximum value $\gamma(k_0)$ can be positive at $Z_n > 2(\Lambda k_0)^2$ due to the nonlinear dependence of the ionization rate on the electron density.

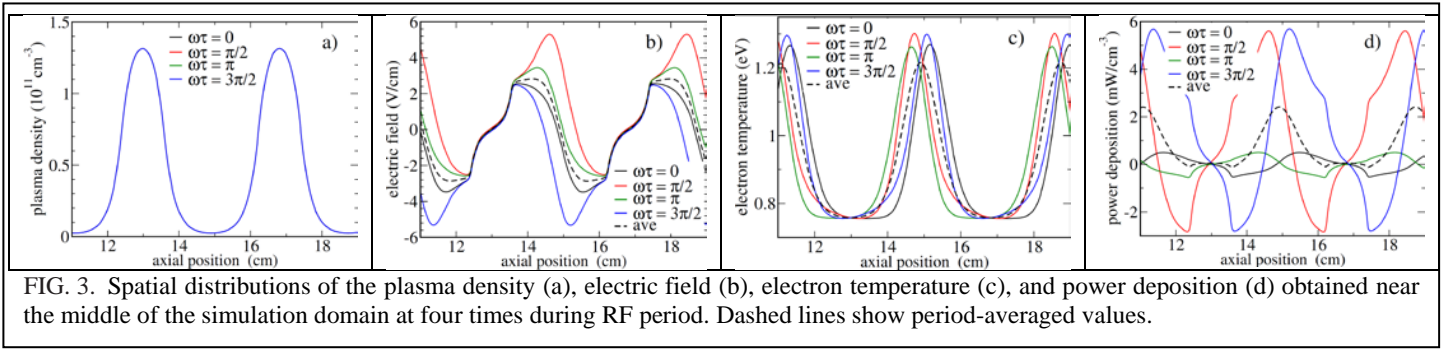

FIG. 3. Spatial distributions of the plasma density (a), electric field (b), electron temperature (c), and power deposition (d) obtained near the middle of the simulation domain at four times during RF period. Dashed lines show period-averaged values.

Under our conditions, the required nonlinearity is due to the stepwise ionization and the Maxwellization of electron distribution function (EDF) due to Coulomb collisions. For calculation of the ionization frequency, we used a simple approximation instead of solving the local Boltzmann equation (as in Ref. [14]):

$$v_i(n,T_e) = v_0 \exp(-\varepsilon_0/T_e) \begin{cases} \exp(n/n_c), & n < n_1 \\ \exp(n_1/n_c), & n > n_1 \end{cases} \quad (5)$$

where $n_c$ controls the rate of the non-linear dependence, and $n_1$ defines the saturation value. The dependence of the ionization frequency on electron temperature $T_e$ is assumed in the Arrhenius form with a parameter $\varepsilon_0$ corresponding to Argon gas.

For numerical simulations, we solved the Poisson equation for the electric field, and could describe the entire structure of glow discharges including near-electrode phenomena. The charged particle densities were calculated by solving the drift-diffusions equations for electrons and ions. We also solved the electron energy transport equation to account for the thermal conductivity of electrons and the related ionization non-locality. The initial conditions correspond to radially and axially uniform plasma column with $T_e = 1$ eV and $n_e = 10^{14}$ m$^{-3}$. For the boundary conditions, see Ref. [14].

First, the computational model was used for one-dimensional (1D) simulations similar to the analysis described above. These simulations were performed for a tube with radius R=1.1 cm and length L=30 cm, for gas pressure 0.5 Torr, over a wide range of driving frequencies (from 0.5 MHz to 100 MHz). By changing the amplitude of the driving voltage, we have observed that the plasma stratification occurs when the mean plasma density in the positive column is within the range defined by Eq. (5). Figure 3 shows an example of the simulation results for 2 MHz, voltage 1000 V, and $n_c = 8\times 10^{15}$ m$^{-3}$.

It is seen from Fig. 3(a) that the plasma density does not change during the RF period. The electric field (see Fig. 3(b)) changes its sign and has zero values at the positions where the plasma density reaches maximal and minimal values. The ambipolar component of the field dominates, while the conductive component results in oscillations of the electric field in time with respect to the average value (dashed line). The electron temperature (Fig. 3(c)) exhibits substantial oscillations during the RF period with maximal values near the minimum of the electron density. The observed oscillations of the electron temperature at steady plasma density correspond to the *dynamic regime of plasma operation* [15]. This regime exists due to the distinction of the plasma ambipolar diffusion time and the free electron diffusion time.

The observed oscillations of the electron temperature are due to the temporal and spatial oscillations of the power deposition shown in Fig.3(d). The time-averaged value of the power deposition is positive and has maximal value at the position of the maximum of the electron temperature. However, during RF period, the power deposition reaches negative values, when the electron current flows against the electric field. Such negative power dissipation was previously observed in the skin layer of ICPs at low gas pressures under conditions of the anomalous skin effect [16].

By changing the driving frequency, we have observed that the temporal oscillations of the electron temperature gradually disappear at frequencies above 30 MHz. With decreasing driving frequency, the amplitude of temperature oscillations increases, and oscillations of the plasma density become noticeable at frequencies below 0.5 MHz.

The spatio-temporal dynamics of the electric field and power dissipation can be well understood using a simple model. The electric field contains two components:

$$E(x,t) = \frac{j(t)}{eb_e n_e(x)} - \frac{T_e}{en_e}\frac{\partial n}{\partial x} = E_j(x,t) + E_a(x) \quad (6)$$

The first component, $E_j(x,t)$, which is due to the conduction current, oscillates in time and space. The second, $E_a(x)$, which is due to the ambipolar field, does not oscillate in time because the plasma density remains steady according to Fig 3. By approximating the axial distribution of the plasma density in the form $n(x) = n_0 + n_1 \cos(kx)$ and introducing dimensionless variables $\tilde{n} = n/n_0$, $\tilde{E} = E/E_0$, where $j = eb_e n_0 E_0 \sin(\omega t)$, we obtain

$$\tilde{E}(\tau,\xi) = (\sin\tau + \alpha A \sin(2\pi\xi))/\tilde{n}(\xi) \quad (7)$$

where $\alpha = n_1/n_0$, $A = \lambda_T/\lambda$, $\tilde{n} = 1+\alpha\cos(2\pi\xi)$, $\tau = \omega t$ and $\xi = x/\lambda$. The distribution of the electric field is fully defined by the ratio of the electron energy relaxation length to the striation length and by the amplitude of plasma density modulation. The distributions of the electric field and power deposition $P = jE$ calculated according to Eq. (7)

for $\alpha = 0.8$ and $A = 3$ are shown in Fig. 4. The negative values of power dissipation, which become noticeable at $\alpha > 1$, are clearly visible in Fig. 4.

For 2D simulations, we employed the recently developed fully-coupled multi-fluid implicit plasma solver with a dynamically adaptive grid, which has already been applied to simulations of moving striations in DC discharges [17]. Symmetric powering of the electrodes with an external circuit with two resistors was used to control the plasma density. The electric potential at the dielectric wall was calculated from the local surface charge, which evolved in time based on the electron and ion fluxes to this wall. Similar to 1D simulations, we have observed that standing striations exist only in a certain range of plasma densities where the non-linear dependence of the ionization rate was substantial ($n_e$ close to $n_C$ in Eq. (5)).

Figure 5 shows an example of 2D simulations for radius $R = 1$ cm, length $L = 14$ cm, gas pressure 0.5 Torr, driving frequency of 20 MHz, and $n_C = 2 \times 10^{15}$ m$^{-3}$. A movie available on the journal web site illustrates that standing striations are formed during a few thousands of RF cycles. It is seen in Figure 5 that in stratified plasma under these conditions, despite substantial time variations of the power deposition (c), not only the electron density (a) but also the electron temperature (b) does not vary in time, except the oscillating sheaths near the electrodes.

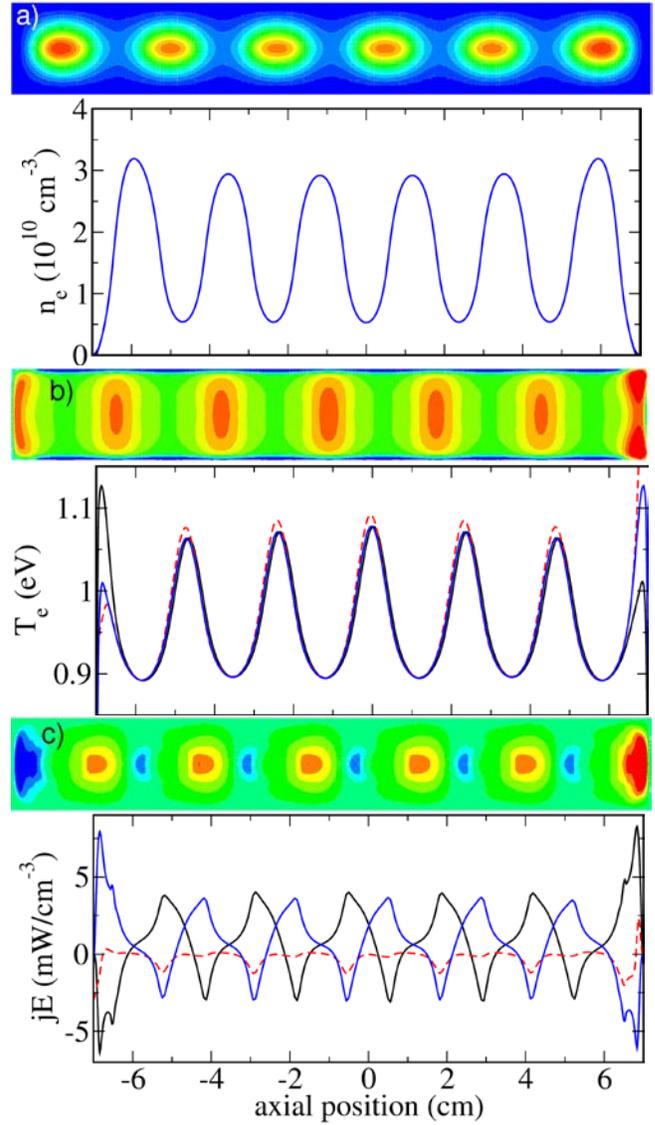

FIG. 5. Instantaneous spatial distributions of the plasma density (a), electron temperature (b) and power dissipation (c). The axial distributions are shown at three times during RF period ($\tau = \pi/2$ (black), $\pi$ (red), and $3\pi/2$ (blue) lines).

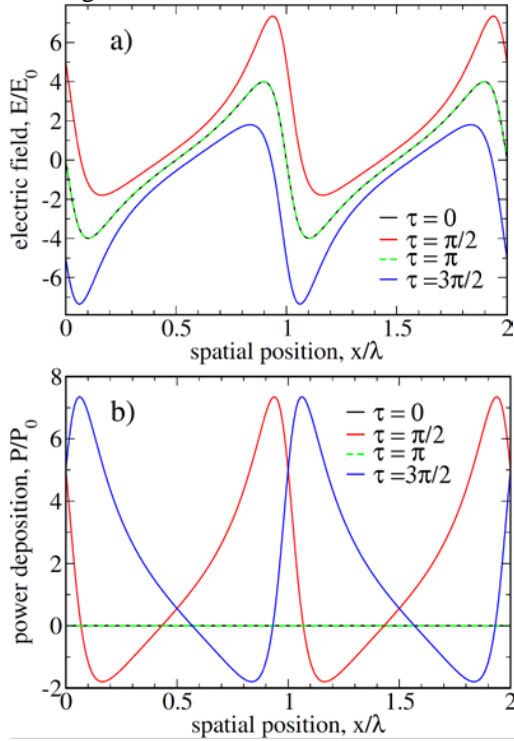

FIG. 4. Instantaneous distributions of the electric field (a) and power dissipation (b) calculated according to Eq. (7).

In summary, we have clarified the nature of standing striations in RF discharges of Argon gas. Experimental studies have confirmed that standing striations exist in Capacitively Coupled Plasma over a range of gas pressures, driving frequencies and plasma densities (driving voltage). Numerical simulations revealed similar nature of standing striations in CCP and moving striations in DC discharges near the Pupp boundary. We confirmed that the non-linear dependence of the ionization rate on electron density is the common underlying mechanism of both phenomena.

Based on our studies, we can predict that the previously observed standing striations in high-frequency discharges of noble gases can be explained by the proposed mechanism. In the general case, the nonlinear dependence of the ionization rate on plasma density could appear from the Maxwellization of the EDF, stepwise ionization, and gas heating. From our previous work, we know that the EDF Maxwellization is strongest in Argon and decreases in strength for Neon and Helium (commonly used noble gases). This may explain why standing striations have been often observed in high-frequency discharges in Argon and rarely seen in Helium. A multi-fluid plasma model could be appropriate for the description of these striations due to frequent Coulomb collisions among electrons under conditions of interest, but the kinetic effects associated with the temporal and spatial non-locality of the EDF may be required to explain the frequency range where these striations exist. For molecular gases, excitation of molecular vibrations and specifics of ionization processes may hide the proposed mechanism and cause new instability channels. The developed computational tools could be further advanced by implementing detailed ionization and recombination mechanisms for quantitative descriptions of plasma stratification in DC and high-frequency discharges.

Thanks to B. Alexandrovich for the help with experiments. This work was partially supported by the NSF EPSCoR project OIA-1655280.

*Corresponding author.
Vladimir.Kolobov@uah.edu